\documentclass[10pt,twocolumn,pra,
 superscriptaddress, longbibliography, 
 aps,showkeys,floatfix]{revtex4-2}

\usepackage{setspace}
\usepackage{enumitem}
\usepackage{ dsfont }
\usepackage[pdftex]{hyperref,color,graphicx}
\usepackage{amsfonts,amssymb,amsmath}
\usepackage[separate-uncertainty=false]{siunitx}
\usepackage[english]{babel}
\usepackage[utf8]{inputenc}
\usepackage[T1]{fontenc}
\usepackage[normalem]{ulem}
\usepackage[toc,page]{appendix}
\usepackage[space]{grffile}
\usepackage{placeins}
\usepackage{latexsym}
\usepackage{textcomp}
\usepackage{longtable}
\usepackage{multirow,booktabs}
\usepackage{url}
\hypersetup{colorlinks=false,pdfborder={0 0 0}}
\usepackage{dcolumn}
\usepackage{bm}
\usepackage{dsfont}
\usepackage{upgreek}
\usepackage{cancel}
\usepackage{braket}
\usepackage{xcolor}
 \usepackage{tikz}

\newcommand{\bbZ}{\mathbb{Z}}

\newcommand{\zZ}{\mathbb{Z}}

\newcommand{\one}{\mathds{1}}

\newcommand{\edit}[1]{{\color{black}#1}}

\begin{document}

\title{Impact of Boundary Conditions on the Double-Kicked Quantum Rotor}

\author{Victoria Motsch}
\email{Contact author: victoria.motsch@stud.uni-heidelberg.de}
\affiliation{Institute of Theoretical Physics, Heidelberg University, Philosophenweg 16, 69120 Heidelberg, Germany}

\author{Nikolai Bolik}
\affiliation{Institute of Computer Science, Heidelberg University, Im Neuenheimer Feld 205, 69120 Heidelberg, Germany}

\author{Sandro Wimberger}
\email{Contact author: sandromarcel.wimberger@unipr.it}
\affiliation{Dipartimento di Scienze Matematiche, Fisiche e Informatiche, Universit\`{a} di Parma, Parco Area delle Scienze 7/A, 43124 Parma, Italy}
\affiliation{INFN, Sezione di Milano Bicocca, Gruppo Collegato di Parma, Parco Area delle Scienze 7/A, 43124 Parma, Italy}

\date{\today}

\begin{abstract}
We study the on-resonance Spin-1/2 Double Kicked Rotor, a periodically driven quantum system that hosts topological phases. Motivated by experimental constraints, we analyze the effects of open and periodic boundary conditions in contrast to the idealized case of infinite momentum space. As a bulk probe for topological invariants, we focus on the Mean Chiral Displacement (MCD) and show that it exhibits a pronounced sensitivity to boundary conditions, which can be traced to the dynamics in momentum space. Under open boundaries, states that would otherwise extend freely become localized at the edges of the finite momentum space, forming quasienergy edge states. While the bulk response measured by the MCD is strongly affected once the evolving wave packet reaches the boundaries, the persistence of these edge states still reflects the bulk-edge correspondence and provides reliable signatures of topological transitions.
\end{abstract}


\maketitle

\section{Introduction}
\label{sec:Intro}

Kicked Rotor Models have long formed the theoretical basis for the exploration of classical and quantum chaos \cite{Casati1979, Izrailev, wimberger2023nonlinear}. The classical Kicked Rotor, described by the Chirikov-Taylor map \cite{Chirikov:2008}, has been used extensively for studying the transition from regular to chaotic dynamics. Its quantum counterpart, the Quantum Kicked Rotor (QKR), has been equally influential, playing a central role in the development of quantum chaos and in the discovery of dynamical localization, an analog of Anderson localization in momentum space \cite{Casati1979, Izrailev, Chirikov_Izrailev_Quantum_Chaos, fishmanDL}.

In 2008, Wang and Gong \cite{PhysRevA.77.031405} discovered that the double-kicked quantum rotor supports a Hofstadter-butterfly-like Floquet spectrum \cite{PhysRevB.14.2239}. Since then, periodically driven kicked rotor systems have been actively explored as toy models for topological phases of matter \cite{PhysRevLett.109.010601, Wang2013KickedHarper, PhysRevB.90.195419}.

One of the key features of the QKR is that it naturally implements discrete-time dynamics: the evolution of the system is defined from kick to kick. Such dynamics can be engineered in different physical contexts. A widely studied framework is that of Quantum Walks, which provide discrete-time analogs of quantum transport  \cite{Kitagawa, Kitagawa_2012}. Alternatively, periodic driving can be used to manipulate the band topology and realize Floquet topological phases in a variety of quantum systems \cite{PhysRevB.79.081406, Lindner2011, Goldman2014, Goldman2016, RevModPhys.89.011004, Zhang02102018}. In particular, it has been proposed that Bose-Einstein condensates, consisting of $^{87}$Rb atoms with internal spin-$1/2$, provide an experimental platform to realize such Floquet systems via sequences of spin-dependent optical lattice pulses \cite{Summy_Wimberger_Rubidium, Gresch_Wimberger_Rubidium, PhysRevA.99.043617, PhysRevResearch.3.043062}. Inspired by this, Zhou and Gong demonstrated that the Spin-$1/2$ Double-Kicked Quantum Rotor (DKQR) hosts Floquet topological phases \cite{Zhou_Gong}. More recently, it was shown that the DKQR is capable of realizing all topologically nontrivial Altland-Zirnbauer (AZ) symmetry classes \cite{PhysRevB.55.1142, PhysRevB.96.155118, PhysRevB.96.195303} in one dimension \cite{Nagoya}. The AZ classes provide a comprehensive classification of quantum systems according to the presence or absence of time-reversal, particle-hole, and chiral symmetries, thereby encompassing all possible symmetry-protected topological phases \cite{PhysRevB.96.195303}. Their realization in a controllable Floquet setting is of particular interest, as periodically driven systems can exhibit richer topological structures than their static counterparts, including exotic phases characterized by $0$- and $\pi$-quasienergy edge states \cite{Wintersperger2020AnomalousFloquet, Zhou_Gong, Gong2025}. These edge states are gapless boundary states that are immune to disorder \cite{PhysRevB.96.195303} and play an important role in the analysis of the Floquet topology of the DKQR \cite{Zhou_Gong, Nagoya, Bolik043318, Groiseau2019}.

The DKQR extends the standard QKR by introducing an internal spin degree of freedom \cite{Scharf1989} and by applying two kicks within each driving period. At resonance, i.e., when the free evolution between kicks becomes the identity, the effective dynamics reduce to a tight-binding Hamiltonian in momentum space \cite{Nagoya, PhysRevA.99.043617, wimberger2023nonlinear}. Combined with the spin structure, this leads to nontrivial topological band structures that can be characterized within Floquet analysis \cite{Zhou_Gong}.

To probe these topological phases of the DKQR, the Mean Chiral Displacement (MCD) has emerged as a meaningful experimental observable: it converges to half the winding number, which serves as the bulk topological invariant of the DKQR \cite{Zhou_Gong, Nagoya, Bolik043318, Groiseau2019}. Theoretical analysis, in particular numerical simulations with an inevitable finite-size cutoff, typically assume either periodic boundary conditions (PBCs) or open boundary conditions (OBCs). In contrast, cold-atom experiments effectively approximate unbounded systems in momentum, at least for the number of kicks typically applied \cite{Summy_Wimberger_Rubidium, Gresch_Wimberger_Rubidium, DA_Sandro}. Only in the limit of non-vanishing kick duration, effective barriers are created in momentum space which are, however, relatively smooth \cite{Blumel_Fishman_Smilanzky} and quite different from the boundary conditions to be studied here \cite{Zhou_Gong, Bolik043318, SADGROVE2011315}.

This raises the central question addressed in this work: How do boundary conditions influence the Floquet spectrum? Also how does the experimentally accessible observable, the MCD, which serves as a proxy for the system's topological invariants, change under the influence of different boundary conditions? Investigating the Floquet spectrum of the on-resonant DKQR confirms the bulk-edge correspondence and shows that the edge states localize at the edge of the momentum basis. The analysis of the MCD reveals a strong dependence on the boundary condition at large kicking strengths. The deviation of the MCD can be explained in terms of the probability distribution in momentum space and the mean momentum.

The remainder of this paper is organized as follows. In Sec. \ref{sec: Background}, we introduce the QKR and DKQR models and derive their Floquet operators. Sec. \ref{sec: BCs} discusses the boundary conditions (BCs) considered. Our results are presented in Sec. \ref{sec: Numerical analysis}: Sec. \ref{sec: Floquet Analysis} analyzes the Floquet spectra, while Sec. \ref{sec: MCD} focuses on the boundary dependence of the MCD. Finally, Sec. \ref{sec: Conclusion} summarizes our findings.

\section{Background} \label{sec: Background}

\subsection{Quantum Kicked Rotor Models} \label{sec: QKR Models}
\subsubsection{Quantum Kicked Rotor}

The simplest quantum-mechanical kicked rotor model is the Quantum Kicked Rotor (QKR). It is the first and standard model used to understand the transition from regular to chaotic motion as well as from classical to quantum behavior \cite{Casati1979, Izrailev, Chirikov_Izrailev_Quantum_Chaos, Nonlinearity, wimberger2023nonlinear, Shepelyansky_Rev} and has been studied extensively over the past decades, see the reviews \cite{Izrailev, Rev_Santhanam, fishmanDL, SADGROVE2011315, Raizen_Finite_Kick}. The dynamics of the QKR are governed by the Hamiltonian \cite{wimberger2023nonlinear}
\begin{equation}
    \hat{H} = \frac{\hat{n}^2}{2} + k \cos(\hat{\theta}) \sum_{m = 0}^\infty \delta(t - m \tau).
    \label{eq: Hamiltonian QKR}
\end{equation}
Here, $\hat{n}$ and $\hat{\theta}$ are the (angular) momentum and position (angle) operators. The angle is confined to $\theta = x \mod 2\pi$, which restricts the system to a finite interval in position space. By Fourier duality, this compactification enforces a discretization in momentum space: the momenta take only integer values $n \in \{\ket{n}| n \in \zZ\}$ in the chosen units  \cite{SADGROVE2011315}. This is directly analogous to Bloch's theorem \cite{bloch}, where imposing periodicity in real space quantizes crystal momentum within the first Brillouin zone. The parameters $k$ and $\tau$ correspond to the potential kicking strength and the time interval in between two successive kicks over which the system evolves freely, respectively.

In the experimental realization \cite{GSZ1992, Raizen_Finite_Kick, SADGROVE2011315}, atoms are subjected to spatially periodic optical lattice potentials. This periodicity allows one to invoke Bloch's theorem \cite{bloch}, which states that momentum separates into an integer part $n$ mentioned above and a quasimomentum $\beta \in [0, 1): p = n + \beta$ \cite{DA_Sandro}. In free space, translational symmetry leads to momentum conservation. Inside a discrete lattice, only discrete translation remains \cite{ashcroft_mermin_1976}. The corresponding conserved quantum number to this discrete symmetry is the quasimomentum. It determines the band dispersion, scattering, and transport, like momentum does in free space. Unlike real momentum, however, it is conserved only modulo a reciprocal lattice vector, reflecting the underlying periodicity of the optical lattice \cite{ashcroft_mermin_1976}. In the following, we restrict ourselves to the case of zero quasimomentum ($\beta = 0$). \edit{Physically, this corresponds to the assumption of zero temperature, as well as no spontaneous emission, which could change the quasimomentum during the temporal evolution.}

The QKR exhibits not only spatial but also temporal periodicity, since $\hat{H}(t) = \hat{H}(t+1)$. This allows the use of Floquet theory to analyze the quasienergy spectrum \cite{Shirley1965, DA_Sandro, Izrailev}. Floquet theory states that a time-periodic Hamiltonian, $H(t+ T) = H(t)$ with period $T$ can be solved by a Schr\"odinger equation of the form
\begin{equation}
    \ket{\psi(t)} = \sum_\epsilon c_\epsilon e^{-i \epsilon t} \ket{\epsilon(t)},
    \label{eq: SEQ Floquet}
\end{equation}
with time-periodic eigenstates $\ket{\epsilon(t)}$ and time-independent expansion coefficients $c_\epsilon \in \mathbb{C}$ \cite{Shirley1965}. The eigenstates $\ket{\epsilon (t)}$ and the corresponding eigenvalues $\epsilon$ solve the stationary eigenvalue problem
\begin{equation}
    \hat{\mathcal{H}} \ket{\epsilon(t)} = \epsilon \ket{\epsilon(t)},
\end{equation}
with the Floquet Hamiltonian $\hat{\mathcal{H}} = \hat{H}(t) - i \partial_t$. The eigenvalues $\epsilon$ are the quasienergies \cite{Shirley1965, Izrailev}. In this work, we follow the convention of referring to a set of quasienergies $\{\epsilon_n\}$, plotted as a function of some control parameter, as quasienergy bands. While for fixed $\beta = 0$, these are discrete eigenvalues of the Floquet operator, their continuous dependence on system parameters such as the kicking strength justifies the band terminology commonly used in the literature \cite{Izrailev, Zhou_Gong}. Based on this, an arbitrary Floquet operator can be written in spectral decomposition
\begin{equation}
    \hat{U} = \sum_n e^{-i \epsilon_n \tau}\ket{n} \bra{n},
\end{equation}
where $\tau$ is the driving period \cite{DA_Sandro}.

For the QKR, the one-cycle Floquet operator factorizes into a free evolution part, $\hat{U}_\tau$, and a kicking part $\hat{U}_k$
\begin{equation}
    \hat{U} = \hat{U}_\tau \cdot \hat{U}_k = e^{-i \frac{\hat{n}^2}{2} \tau} \cdot e^{-i k \cos(\hat{\theta})}.
    \label{eq: Floquet Operator QKR}
\end{equation}

Two quantum regimes are particularly relevant \cite{Izrailev}:
\begin{itemize}
    \item Resonance: $\tau = 4\pi$. Here, the free evolution term reduces to identity, $\hat{U}_\tau= \one$, so the Floquet operator is governed entirely by the kicking term $\hat{U}_k= e^{-i k \cos(\theta)}$. In this case, an analytic solution exists \cite{Izrailev}: the quasienergy spectrum is
    \begin{equation}
    \label{eq:FS-QR}
        \epsilon = \frac{k \cos(\theta)}{4\pi},
    \end{equation}
    which is continuous and has a cosine form. Dynamically, the system exhibits ballistic transport in momentum space, with quadratic energy growth, $E(t) = \bra{\psi(t, \theta)}-\frac{1}{2} \frac{\partial^2}{\partial \theta^2} \ket{\psi(t, \theta)} \propto t^2$ \cite{Izrailev}. The quantum resonant regime has been useful for a plethora of applications of the QKR, see, e.g., Refs. \cite{SADGROVE2011315, Gong2025, Gresch_Wimberger_Rubidium, Summy2010, Sadgrove2007, LZG2008, Ratchet2017, WRH2013, Kanem2007, Ryu2006, Andersen2018, gil2008, Weiss2015, WTP2009, Andersen2020}.
    \item Antiresonance: $\tau = 2\pi$. Two successive kicks cancel each other, so that $\hat{U}_{\rm free} = \pm \one$. As a result, the wavefunction becomes periodic with a period of two kicks, and the quasienergies reduce to $\epsilon = 0, \pm \pi$, which are highly degenerate \cite{Izrailev}. The quasienergy bands are completely flat, leading to suppressed transport \cite{PhysRevE.54.5948, Izrailev, Summy2010}.
\end{itemize}

\subsubsection{Double-Kicked Quantum Rotor (DKQR)}

Introducing an additional internal spin-1/2 degree of freedom, and kicking the rotor not only once, but twice per cycle with different kicking strengths $k_1$ and $k_2$, leads to the so-called Double Kicked Quantum Rotor (DKQR). This model is of special interest because it can host topological phases \cite{Kitagawa, Zhou_Gong, Nagoya}. The definition and classification of such phases rely on the presence and absence of certain symmetries. In particular, we consider the framework of the tenfold-way classification, which distinguishes systems according to time-reversal symmetry, particle-hole symmetry, and sublattice or ''chiral'' symmetry (see \cite{PhysRevB.96.195303} for the definitions). These three symmetries give rise to the ten AZ classes \cite{RevModPhys.88.035005, Ryu2010_NJP}. It was recently shown  that the DKQR alone can realize all topologically nontrivial AZ classes \cite{Nagoya}, highlighting its potential as a promising platform for simulating topological phenomena in condensed matter physics.

The Hamiltonian of the DKQR can be written in a general form as \cite{Zhou_Gong}
\begin{equation}
    \begin{split}
        \hat{H} &= \frac{\hat{n}^2 \otimes \one}{2} 
        + k_1 \cos(\hat{\theta}) \otimes 2 \hat{S_x} \cdot \sum_{m = 0}^\infty \delta(t - m T) \\
        &\quad + k_2 \sin(\hat{\theta}) \otimes 2 \hat{S_y} \cdot \sum_{m = 0}^\infty \delta(t - mT - \tau_1),
    \end{split}
    \label{eq:dkqr-ham}
\end{equation}
where $T = \tau_1 + \tau_2$. Here, $\tau_1$ and $\tau_2$ denote the free-evolution intervals following the kicks of strengths $k_1$ and $k_2$, respectively. $\hat S_x$ and $\hat S_y$ are the spin operators acting on the internal spin-1/2 space of the rotor. In Pauli matrix representation we have $\hat S_i \doteq \frac{\hbar}{2} \sigma_i$, with $\sigma_i, \,  i \in \{x, y\}$ and  $\hbar=1$ in our chosen units. Practically, the sin-lattice can be realized by phase shifting the cos-lattice by $\frac{\pi}{2}$ at each second kick \cite{Zhou_Gong, Groiseau2019, BA_Nikolai}.
Eq. \eqref{eq:dkqr-ham} induces the Floquet operator \cite{Zhou_Gong}
\begin{equation}
    \hat{U}_{\rm DKQR} = e^{-i \frac{\hat{n}^2}{2} \tau_2} e^{-i k_2 \sin(\hat{\theta}) \sigma_y} e^{-i \frac{\hat{n}^2}{2} \tau_1} e^{-i k_1 \cos(\hat{\theta}) \sigma_x}.
    \label{eq: Floquet operator DKRS}
\end{equation}

To define topological phases, phases, the authors of Refs.\cite{Kitagawa, Zhou_Gong} argued that the Floquet operator $\hat{U}_{\rm DKQR}$ has to be transformed into a chiral symmetric time frame. Moreover, we solely focus on the resonant case here, $\tau_1 = \tau_2 = 4\pi$. The resulting Floquet operator used for our calculations \edit{then is}
\begin{equation}
    \hat{U}_1 = e^{-i \frac{k_1}{2} \cos(\hat{\theta}) \sigma_x} e^{-i k_2  \sin(\hat{\theta})\sigma_y} e^{-i \frac{k_1}{2} \cos(\hat{\theta}) \sigma_x},
    \label{eq: Chiral symmetric Floquet operator DKRS}
\end{equation}
\edit{which symmetrizes} the $k_1$ kick around the $k_2$ kick, in accordance with \cite{Zhou_Gong}. Since the Floquet operators in different chiral symmetric frames are related via unitary transformations, they share a common quasienergy spectrum, such that we can focus on studying it in this particular frame.

\subsubsection{Bulk-Edge Correspondence}

The bulk-edge correspondence, as explained, e.g., in Refs. \cite{PhysRevB.86.195414, Zhou_Gong} establishes a direct link between the bulk topology and the existence of edge states. In detail, it tells us that the absolute value of the winding number $W_0$ ($W_\pi$) equals the number of degenerate edge-state pairs at zero ($\pi$) quasienergy. This correspondence is of particular interest because it translates an abstract topological invariant into a quantity that is experimentally more easily accessible.
Moreover, in the DKQR the winding numbers can be tuned directly by varying the kicking strengths $k_i$, enabling access, in principle, to regimes with large winding numbers and correspondingly many edge states. Such tunability provides a platform to study exotic transport phenomena and robustness properties of topological phases in a highly controllable setting \cite{Zhou_Gong}. As the DKQR can realize all nontrivial AZ classes in 1D \cite{Nagoya}, the DKQR offers a minimal yet versatile model for simulating the diverse topological phenomena otherwise realized in very different physical systems, e.g., one-dimensional topological superconductors \cite{Kitaev2001}, chiral symmetric insulators \cite{Su1979} or spinless superconducting wires \cite{Schnyder2008}.

The winding numbers can be calculated as in Refs. \cite{maffei2018, cardano2017}
\begin{align}
\begin{aligned}
    W_l &= \int_{-\pi}^\pi \frac{d\theta}{2\pi} (\vec{n}_l \times \partial_\theta \vec{n}_l)_z,\quad (l = 1,2)
    \label{eq: winding number}\\
    W_0 &= \frac{W_1 + W_2}{2}, \quad W_\pi = \frac{W_1 - W_2}{2},
\end{aligned}
\end{align}
and are the topological invariants by which the topological phases of the DKQR are defined \cite{Zhou_Gong}.

\subsubsection{Mean chiral displacement (MCD)}

The proposed experimentally accessible observable for the DKQR is the Mean Chiral Displacement (MCD) \cite{Zhou_Gong, Bolik043318, Groiseau2019}. Formally, the MCD is the expectation value of the chiral displacement operator $\hat{C}(t) = \hat{U}^\dagger (t) (\hat{n} \otimes \hat{\Gamma} )\hat{U}(t)$ after $t$ unitary time evolutions $\hat{U}(t)$. In the on-resonant DKQR, the chiral symmetry operator is given by $\hat{\Gamma} = \sigma_z$ such that the chiral displacement of the system, prepared in the initial state $\ket{\psi_0}$ after $t$ evolution steps under the Floquet operator $\hat{U}(t)$ is given by
\begin{equation}
    C(t) = \bra{\psi_0} \hat{U}^{-t}(\hat{n} \otimes \sigma_z) \hat{U}^t\ket{\psi_0}.
    \label{eq: MCD definition}
\end{equation}
The chiral displacement $C(t)$ depends on the formulation of the Floquet operator $\hat{U}(t)$. We use the chiral symmetric formulation of the Floquet operator given in Eq. \eqref{eq: Floquet operator DKRS}, such that  $C_1(t) = \bra{\psi_0} \hat{U}_1^{-t} (\hat{n} \otimes \sigma_z) \hat{U}_1^t \ket{\psi_0}$.

The chiral displacement is additionally averaged over $t$ driving periods, resulting in the MCD. It has been shown that this quantity converges to half the winding number $W_l \ (l = 1,2)$, Eq. (\ref{eq: winding number}) \cite{cardano2017, Zhou_Gong}. The MCD is given by
\begin{align}
\begin{aligned}
    \overline{C_l(t)} &= \frac{1}{t} \sum_{t'=1}^t C_l(t') \\
    &= \frac{W_l}{2} - \int_{-\pi}^\pi \frac{d\theta}{2\pi} \frac{\cos[E(\theta)t]}{2} (\vec{n_l} \times \partial_\theta \vec{n_l})_z \\
    & \overset{t\gg 1}{\longrightarrow} \frac{W_l}{2}.
    \label{eq: Averaged MCD}
\end{aligned}
\end{align}
Practically, the MCD measures the asymmetry between the momentum average of the two spin positions in the $z$ basis. In a perfectly symmetric quantum walk of a single-kicked rotor, the MCD would be zero, see, e.g., \cite{Bolik2024, Bolik2022}. In the on-resonant DKQR, in contrast, as proposed in \cite{Zhou_Gong, Bolik043318}, the evolution is not at all ballistic, and the quantized asymmetry directly reflects the topological phase.

\begin{figure*}[t]
  \centering
    \includegraphics[width=0.45\linewidth]{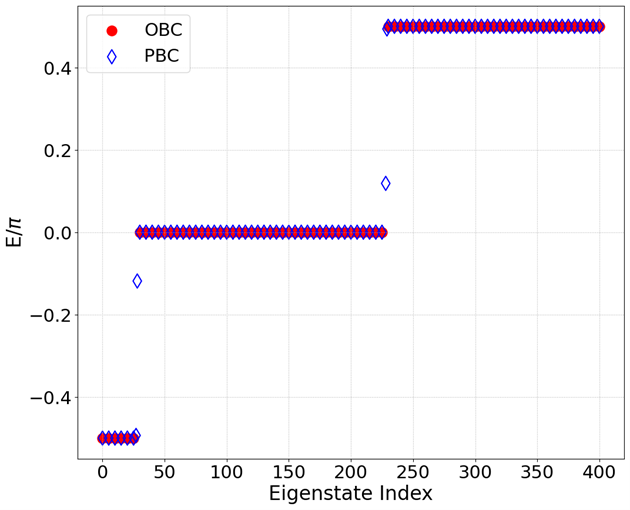}
  \hfill
    \includegraphics[width=0.45\linewidth]{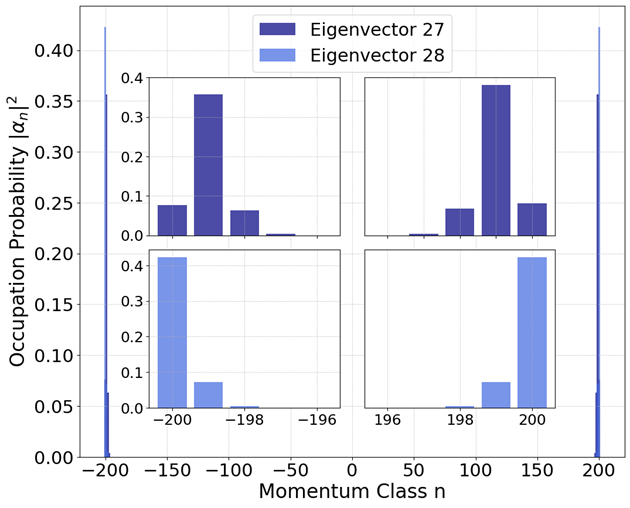}
  \caption{(Left panel) Quasienergy spectrum of the antiresonant QKR under open (red) and periodic (blue) BCs. \edit{The eigenenergies are plotted against their eigenstate index, which is chosen such that the quasienergies are sorted in ascending order.} While OBCs yield flat bands at $\epsilon = 0, \pm 0.5\pi$, PBCs introduce discrete quasienergies that deviate from this flat structure. \edit{For the sake of legibility, only every fifth quasienergy is plotted, as well as all deviating quasienergies.} (Right panel) Occupation probability of the eigenvectors corresponding to the deviating quasienergies under PBCs. Only the eigenvectors with index 27 \edit{(dark blue)} and 28 \edit{(light blue)} are shown; the remaining two exhibit the same edge-localized distribution.}
  \label{fig1}
\end{figure*}\subsection{Boundary Conditions} 
\label{sec: BCs}

We consider three types of boundary conditions (BCs). Since the rotor is restricted to the angular domain $\theta \in [0, 2\pi)$, the conjugate momentum $n$ is quantized and takes only integer values, $n \in \bbZ$, in the units we are using here; see \cite{SADGROVE2011315} for details.

\subsubsection{Open Boundaries}

For the open boundary condition (OBC), momentum space is restricted to a finite set of accessible states. We introduce a maximum momentum that acts as a hard cutoff, such that the rotor can only occupy momentum classes within the interval $n \in [-n_{max}, n_{max}]$. States outside this interval are excluded from the dynamics, i.e., the wave function vanishes at the boundary, $|\psi(n)|^2 = 0$ for $n \notin [-n_{max}, n_{max}]$. OBCs are automatically realized, e.g., in periodically kicked molecules with a fixed number of angular momentum states \cite{Karle_Experimental_Realization, Karle_Edge_States, Karle_Paper}.

\subsubsection{Periodic Boundaries}

In contrast to the open boundary, the periodic boundary condition (PBC) identifies the first and last momentum classes with each other, effectively mapping the momentum space onto a torus. Formally, this is expressed by $\ket{n} = \ket{n + N}$, where $N = 2n_{max} +1$ is the total number of momentum states \cite{Zhou_Gong}. This definition leads to the formation of standing-wave solutions in momentum space.

The numerical implementation of PBCs can be done straightforwardly. For the QKR and DKQR, the Hamiltonian has a band-diagonal form in momentum space, with a single off-diagonal element. In this matrix representation, PBCs are imposed by adding coupling terms to the corner elements on the upper right and lower left, thereby connecting the boundary states.

It is important to point out that this procedure only works at the level of the Hamiltonian, which is Hermitian and band-diagonal. The corresponding Floquet-operator, obtained as the matrix exponential of the Hamiltonian, is unitary and generally has a more complex structure. Therefore, in numerical calculations, we first write the Hamiltonian in matrix form, implement PBCs in that representation, and then compute the Floquet-operator via its matrix exponential. PBCs are often used in numerical simulations for the time evolution of $\delta$-kicked systems based on Fast-Fourier-Transform methods.

\subsubsection{Ideal case of infinite system}

To provide a point of reference and to better isolate the effects of different boundary conditions, we also simulate the ideal case, in which the rotor never interacts with the borders of the momentum basis. This requires choosing a momentum basis that is sufficiently larger than the maximal spread of momentum states reached during the dynamical evolution.

Numerical verification showed that, for the DKQR at $t = 15$ kicks with kicking strengths $k_1 = 1.5\pi$ and $k_2 = 2.5\pi$, the momentum distribution extends up to approximately $n_{\rm max, \,DKQR} \approx 75$ without reaching any boundaries. To safely avoid boundary influences, we therefore set the basis size to $n \in [-200, 200]$. Choosing the basis two to three times larger than the maximal momentum evolution ensures that the dynamics remain unaffected by the boundaries, while still keeping numerical runtimes manageable. What is said for the numerics is true in the same manner for the experimental realization. For kicked cold atoms or Bose-Einstein condensates, boundary effects can be neglected up to some critical kick number, above which the finite duration of the kick may manifest itself. \edit{When the atoms become too fast, they effectively average over the potential, which is periodic in space, and see no more kick. Hence, a finite kick duration instead of a true $\delta$-kick leads to the rise of an effective soft-wall in momentum space. This effect is purely classical, still the wavefunction can partly overcome these barriers by quantum tunneling leading to a rather soft wall. For details on the influence of a finite kicking duration, please see the theoretical and experimental studies reported in \cite{Blumel_Fishman_Smilanzky, SADGROVE2011315, Raizen1998}.}

\section{Impact of boundary conditions on spectrum and MCD} 
\label{sec: Numerical analysis}

\begin{figure*}[tb]
  \centering
    \includegraphics[width=0.32\linewidth]{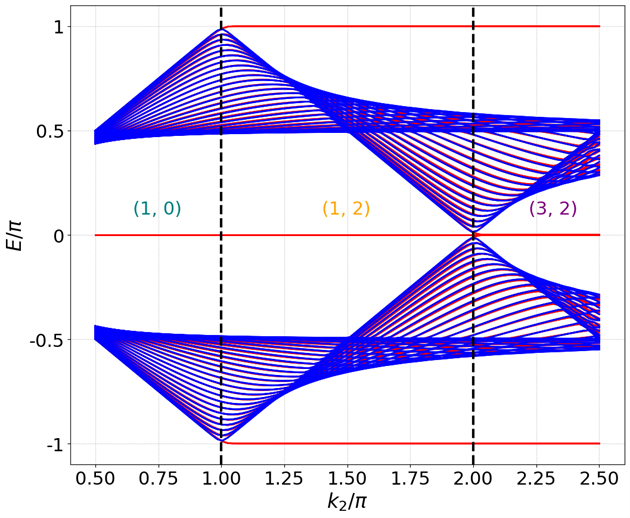}
  \hfill
    \includegraphics[width=0.32\linewidth]{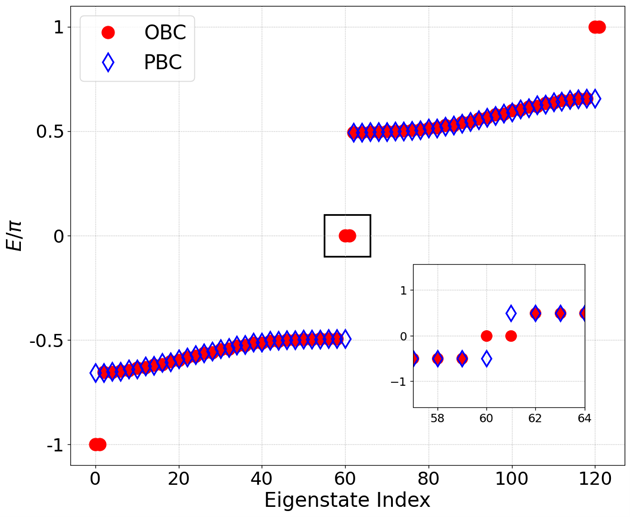}
  \hfill
    \includegraphics[width=0.32\linewidth]{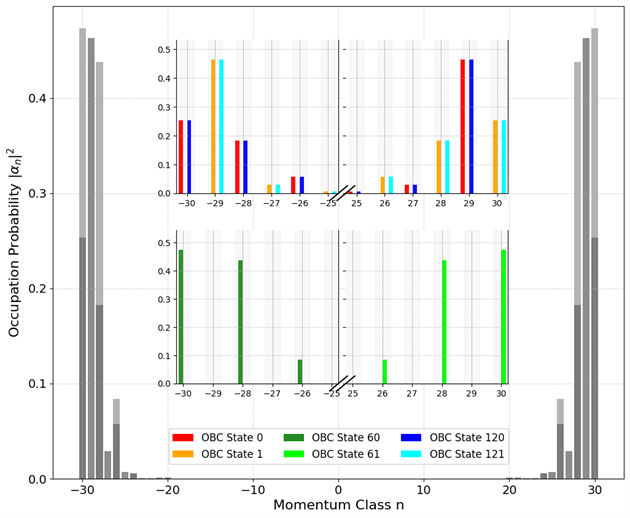}
  \caption{(Left panel) Quasienergy spectrum of the on resonant DKQR as a function of $k_2$ under open (red) and periodic (blue) BCs. Edge states appear at $\epsilon = 0, \pm \pi$ only for OBCs. The pairs of numbers in the areas separated by the black dotted lines are the winding numbers and uniquely define the topological phases. (Middle panel) Localization of the eigenvectors corresponding to the edge states of the quasienergy spectrum at $\epsilon = 0$ and $\epsilon = \pm \pi$. (Right panel) \edit{Edge states appear only under OBCs, and localize at the boundaries of the momentum basis, consistent with the bulk-edge correspondence. The gray bars show the occupation probabilities of all six edge states, illustrating their boundary localization; overlaps occur where several edge states occupy the same momentum class $n$. To resolve these overlaps and analyze the symmetry structure of the edge states more clearly, the insets display the occupation probabilities of each edge state individually in distinct colors, with bars at the same $n$ plotted side-by-side. The insets reveal that the edge states form antisymmetric pairs given by the indices (0,1), (60,61), and (120,121).}}
  \label{fig:2}
\end{figure*}

In the sequel, we analyze the Floquet spectrum of the on-resonant DKQR under different BCs. In particular, the occurrence of edge states will be discussed. In a second step, we investigate how the MCD, as the experimentally accessible quantity, is impacted by the choice of BCs. 

\subsection{Floquet Analysis} 
\label{sec: Floquet Analysis}

\subsubsection{Quantum Kicked Rotor (QKR)}

We start with the simpler QKR to illustrate the effect of the BCs and the occurrence of edge states in momentum space. In the resonant regime, the quasienergy spectrum is the same for PBCs and OBCs, and it agrees with the analytical prediction given in Eq. \eqref{eq:FS-QR}. 
However, investigation of the quasienergy spectrum in the antiresonant regime shows a direct and \edit{subtle} influence exerted by the BCs. 

In the ideal case or for OBCs, the quasienergy bands are completely flat and take values $\epsilon = 0, 0.5\pi$ \cite{Izrailev}. 

Under PBCs, however, discrete points appear at which the quasienergies do not follow the overall flat structure of the band but, in contrast, lie in between those bands, see the left panel in Fig. \ref{fig1}. We investigate the eigenvectors corresponding to these 'deviating' quasienergies. For each of those quasienergies, we plot the occupation probability of all momentum classes, see the right panel in Fig. \ref{fig1}. We found that the corresponding eigenvectors localize at the edge of the momentum basis. This localization is strict, in the sense that only $\approx 4$ momentum classes at each edge are occupied, while the occupation probability for all other momentum classes vanishes. However, we point out that the cause of these 'deviating' quasienergies and their localized eigenvectors is not found in an underlying topological structure. Instead, going from quantum resonant to antiresonant conditions, the effective description of the Floquet matrix doubles in size, see \cite{Izrailev} for details, introducing an effective space doubling, just as the spin 1/2 of a kicked two-level system like our DKQR. However, the standard QKR at antiresonance is non-transporting in momentum space due to the flat bands mentioned above, see \cite{Izrailev}, and topological phases do not manifest. \edit{In other words, we see edge states without any topological meaning. The mentioned doubling arises in antiresonance since even momenta $n=2k$ ($k$ integer) give $+1$ while odd momenta $n=2k+1$ ($k$ integer) give $-1$ in the free rotation part of the Floquet operator, see Eq. \eqref{eq: Floquet Operator QKR}. Consecutive kicks hence cancel each other's effect exactly. All this is well described in the review by Izrailev \cite{Izrailev}. Choosing our box from $[-N,N]$ with PBCs does not respect the corresponding $Z_2$ symmetry of $\pm 1$. If instead we chose a momentum window $[-N-1,N]$, or similar, the spectra of PBCs and OBCs would perfectly coincide (we checked this without showing it here explicitly). The reason is that the shuffling from the upper to the lower end of the box and vice versa introduces a minus sign on the effective value of $n$ on the edge, and a perfect antiresonance is found only when this effect is compensated by a corresponding asymmetric choice $[-N-1,N]$.} 

\subsubsection{On-Resonance DKQR}

Studying the On-Resonance DKQR, where $\tau_1 = 4\pi = \tau_2$, we could confirm the bulk-edge correspondence described in \cite{Zhou_Gong}. This means that edge states arise under OBCs; see Fig. \ref{fig:2}\edit{, left panel}. The number of pairs of degenerate edge states at $\epsilon = 0$ and $\epsilon = \pm \pi$, by the bulk-edge correspondence, equals the winding number, $W_0$ and $W_\pi$, respectively. The pair of these winding numbers, $(W_0, W_\pi)$ uniquely characterizes the topological phase of the system. Plotting the cross section of the quasienergy spectrum at a given kicking strength $k_2$ allows us to count the number of degenerate edge state pairs, see e.g. middle panel of Fig. \ref{fig:2}, where $k_2 = 1.5\pi$.

Again, as in the case of the antiresonant QKR, we investigate the eigenvectors corresponding to these edge states. Plotting the occupation probability, we find that the eigenvectors of the edge states localize strictly at the edges of the momentum basis, see the right panel in Fig. \ref{fig:2}.

It should be noted that the cause for the localization of the edge states in the on-resonant DKQR is the underlying topology, in contrast to the QKR, where we have no underlying topological structure.

Since the spectrum is hardly accessible in kicked atom experiments, just as in solid-state realizations, the MCD as a transport quantity was proposed that reflects the topological phases \cite{Zhou_Gong, Bolik043318, Nagoya}. The next section studies the impact of BCs on the MCD. 

\subsection{Transport - Mean Chiral Displacement} 
\label{sec: MCD}

The MCD serves as a transport observable of the system and is therefore a bulk variable. Strictly speaking, it is defined only in the case of an \edit{infinitely large} system. Since any experiment is practically done on a finite momentum grid, due to the finite kick duration (see \cite{Blumel_Fishman_Smilanzky, DA_Sandro, SADGROVE2011315, Raizen_Finite_Kick}), it is particularly relevant to understand how the MCD is affected by different boundary conditions. \edit{For all our numerical simulations of transport, we chose the initial state $\ket{\psi_0} = \ket{\delta_{n,0}} \otimes \ket{\uparrow}$, with a fixed momentum class $n=0$ and a fixed spin state (spin up).}

\begin{figure}[tb]
    \centering
    \includegraphics[width=\linewidth]{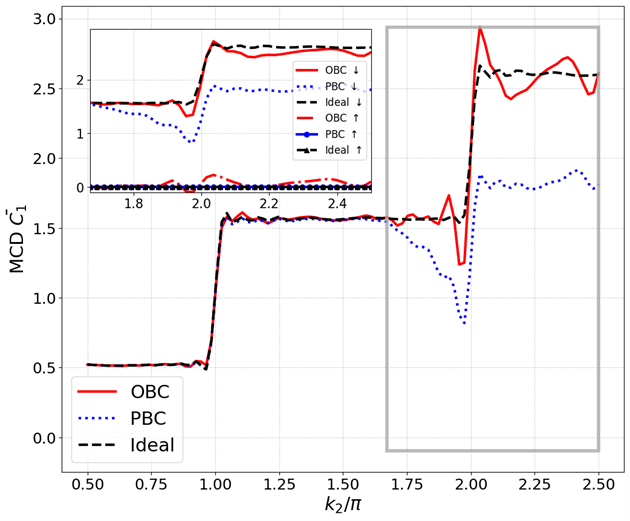}
    \caption{MCD under open (red solid line) and periodic (blue dotted line) boundary conditions, as well as for the ideal case (black dashed line). The inset shows the MCD for the Spin-$\uparrow$- and -$\downarrow$-component separately. We see that the $\downarrow$-component dominates the behavior of the MCD after $k_2 \approx 1.7\pi$. \edit{The initial state is chosen as $\ket{\psi_0} = \ket{\delta_{n,0}} \otimes \ket{\uparrow}$.}}
    \label{fig:3}
\end{figure}

Figure \ref{fig:3} shows the numerically calculated MCD under all three boundary conditions. For strong kicking, $k_2 > 1.7\pi$, the MCD exhibits clear differences depending on the chosen BCs. Compared to the ideal case, PBCs, on the one hand, induce a significant drop of the peak just before the second step arises as a function of $k_2$. This is easily explained since, at larger $k_2$ the evolution is faster in momentum space, and a part of the probability, corresponding to large momenta, will be injected into the other side due to the PBCs. OBCs, on the other hand, lead to reflections of the evolving momentum classes at the boundaries. \edit{This reflection at the boundary, in contrast to PBCs, where $n_{max}$ is identified with $-n_{max}$, does not immediately induce a change of sign of the momenta, but reflects the wavefunction similar to an infinitely high potential wall. Hence, the wavefunction is superposed of an incoming and a reflected part. Therefore, the response of our system is more disturbed.} All that has just been said will depend on the precise parameters, the speed of the evolution, determined by $k_{1,2}$, and the cut-off momentum at which the BCs are defined.

We can decompose the MCD into contributions from the $\ket{\uparrow}$ and $\ket{\downarrow}$ components in the spin $z$ basis. This analysis reveals that only the $\ket{\downarrow}$ component contributes significantly, as the $\ket{\uparrow}$ contribution evolves symmetrically, and by definition, the symmetric part does not contribute to the MCD.

\begin{figure}[tb]
    \centering
    \includegraphics[width=0.95\linewidth]{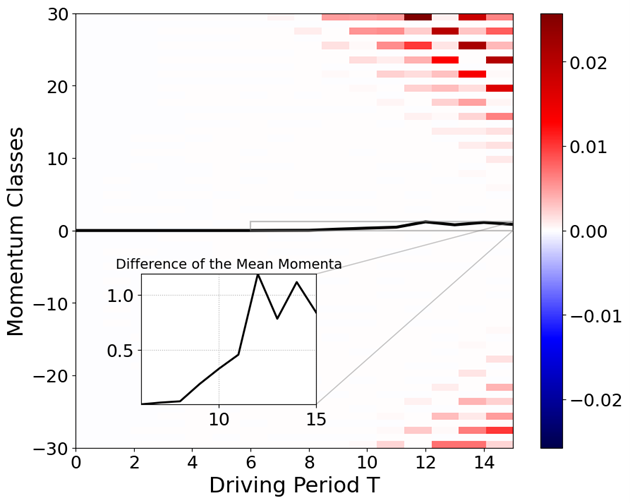}
    \caption{Comparison of the probability distribution of the $\ket{\downarrow}$-component under PBCs minus that of the ideally infinite system. We see a shift of the occupation probability towards positive momenta for PBCs compared to the ideal case. The mean momentum is plotted in black. \edit{Again, the initial state is chosen as $\ket{\psi_0} = \ket{\delta_{n,0}} \otimes \ket{\uparrow}$.} As shown in the inset, after $t \approx 8$ kicks, the mean momentum shifts towards positive values as well. Notice that the spin-down component contributes negatively to the MCD, see its definition in Eq. ~\eqref{eq: MCD definition}, explaining the drop at $k_2 \gtrsim 1.7\pi$ in Fig. \ref{fig:3}.}
    \label{fig:4}
\end{figure}

We further analyze the behavior of the MCD under PBCs by examining the probability distribution and the mean momentum as functions of the momentum classes and the number of driving periods. Figure \ref{fig:4} compares the probability distributions of the on-resonance DKQR in the ideal case and with PBCs. We plot the difference $\Delta |\psi(n, \edit{T})|^2 = |\psi_{\rm PBC}(n, \edit{T})|^2 - |\psi_{\rm Ideal}(n, \edit{T})|^2$ \edit{, where T is the number of successively applied driving periods}. This comparison reveals that, under PBCs, the probability distribution becomes asymmetric: a larger fraction of the probability accumulates at positive momentum classes compared to negative ones. \edit{This can be understood by examining the role of PBCs. First, recall that in the regime of $k_1$ and $k_2$ where the topological phase is non-zero, the DKQR is intrinsically asymmetric. The topological phase, characterized by a pair of winding numbers as shown in \cite{Zhou_Gong}, can be accessed via the MCD. However, by definition, the MCD measures the asymmetry of the evolution. Hence, whenever the MCD (and thus the topological phase) is non-zero, the DKQR dynamics are inherently asymmetric.

On top of this intrinsic asymmetry, the BCs further shape the dynamics. Under PBCs, the identification $\ket{n} = \ket{n+N}$, makes the maximal momentum class $\ket{n_{max}}$ identical to $\ket{-n_{max}}$. As a result, when the wavefunction reaches the lower end of the momentum basis, it reappears at positive momentum. This effectively induces a net shift toward positive momenta. This can be seen in Fig. \ref{fig:4}, where the probability distribution becomes biased towards positive momentum classes once the system has interacted with the periodic boundaries (for driving periods $T \gtrsim 8$). The combined influence of the DKQR's intrinsic asymmetry and its enhancement provided by the choice of PBCs explains the asymmetry of the momentum distribution illustrated in Fig. \ref{fig:4}.

To now understand why the MCD is lowered under PBCs compared to the ideal case shown in Fig. \ref{fig:3}, note that only the $\ket{\downarrow}$ component contributes to the MCD -- and it does so with a negative sign, see its definition in Eq. \eqref{eq: MCD definition}. Consequently, the resulting correction lowers the MCD under PBCs compared to the ideal case.}

A detailed analysis of the MCD reveals that the deviations under both PBCs and OBCs, relative to the ideal case, can be entirely explained by a shift of the mean momentum inherent to the definition of the MCD in Eq. \eqref{eq: MCD definition}. To clarify this, we compare the mean momentum not with the time-averaged MCD (as in \cite{Zhou_Gong, Nagoya}), but with the time-dependent Chiral Displacement $C(\edit{T})$, which explicitly depends on the driving period $\edit{T}$. After accounting for the sign contribution from the $\sigma_z$ operator, the mean momentum is found to coincide with the Chiral Displacement for all boundary conditions, showing that the two quantities differ only by a sign originating from $\sigma_z$ in Eq. \eqref{eq: MCD definition}. Consequently, the deviations of the MCD from the infinite-system limit under different BCs can be attributed entirely to the corresponding momentum-space distribution of the wave function. Moreover, averaging the time-dependent Chiral Displacement $C(\edit{T})$ over the driving period reproduces the MCD values shown in Fig. ~\ref{fig:3}.

Future experiments will have to take into account the effect of the boundaries to measure clear signatures of the topological phases of the DKQR. The boundaries, introduced by experimental artifacts, should also be sufficiently far away to obtain unperturbed experimental MCD data and to mimic the infinite-system case where boundary conditions have no impact.

\section{Conclusion}
\label{sec: Conclusion}

We investigated how finite momentum-space boundaries affect spectrum and transport in the on-resonant DKQR-system. With OBCs we observed $0$- and $\pi$-quasienergy states localized at the momentum edges, consistent with expectations from bulk–edge correspondence in the relevant parameter range. In addition, these edge states are characterized by eigenvectors sharply localized at the momentum boundaries, demonstrating a manifestation of bulk–edge correspondence in this setting.

For transport, the MCD becomes boundary-sensitive once the evolving state reaches the momentum cutoff $n_{\max}$. Under PBCs, the identification of the edge classes transfers population across the cutoff, skews the momentum distribution of the spin-down branch, and lowers the apparent MCD plateau. Under OBCs, reflections at $|n|=n_{\max}$ cause oscillations in the topological observable. We find that $C(\edit{T})$ closely tracks the mean momentum $\langle n\rangle$ (up to a sign), so deviations from the ideal plateau arise from an asymmetric momentum-space evolution of that branch rather than from a change of bulk topology. Crucially, once these boundary-induced asymmetries are modeled or mitigated, for instance, by extending the momentum basis so that the wave packet never reaches the cutoff, or by explicitly accounting for population transfer across the periodic boundary and reflections at open boundaries, the MCD remains a reliable bulk diagnostic of the underlying topological phase for both PBC and OBC.
Earlier DKQR studies \cite{Zhou_Gong, Bolik043318} typically assumed an infinite momentum basis, while we explicitly quantify the effects of a finite momentum basis on the topological observable.

Future work could involve engineering of topological phases in periodically driven Floquet systems \cite{Floquet-2025} with quantum control methods, see e.g. \cite{SB2024}, to enhance the performance of their practical realizations. These could allow, for instance, the preparation of specific Floquet states that characterize a topological phase in state-of-the-art experiments.

\acknowledgements
S.W. acknowledges funding by Q-DYNAMO (EU HORIZON-MSCA-2022-SE-01) with project No. 101131418, and by the National Recovery and Resilience Plan, through PRIN 2022 project "Quantum Atomic Mixtures: Droplets, Topological Structures, and Vortices", project No. 20227JNCWW, CUP D53D23002700006, and through Mission 4 Component 2 Investment 1.3, Call for tender No. 341 of 15/3/2022 of Italian MUR funded by NextGenerationEU, with project No. PE0000023, Concession Decree No. 1564 of 11/10/2022 adopted by MUR, CUP D93C22000940001, Project title "National Quantum Science and Technology Institute" (NQSTI).


%

\end{document}